\documentclass{mn2e}
\usepackage{graphics}
\usepackage{epsfig}
\usepackage{lscape}

\def\beq{\begin{equation}}                          
\def\eeq{\end{equation}}                              
\def\beqa{\begin{eqnarray}}                         
\def\eeqa{\end{eqnarray}}                             
\def\beqan{\begin{eqnarray*}}                      
\def\eeqan{\end{eqnarray*}}                          

\newbox\grsign \setbox\grsign=\hbox{$>$} \newdimen\grdimen \grdimen=\ht\grsign
\newbox\simlessbox \newbox\simgreatbox
\setbox\simgreatbox=\hbox{\raise.5ex\hbox{$>$}\llap
     {\lower.5ex\hbox{$\sim$}}}\ht1=\grdimen\dp1=0pt
\setbox\simlessbox=\hbox{\raise.5ex\hbox{$<$}\llap
     {\lower.5ex\hbox{$\sim$}}}\ht2=\grdimen\dp2=0pt

\title[Infrared-red Cores in Nearby Elliptical Galaxies]
   {Infrared-red Cores in Nearby Elliptical Galaxies}

\author[Yuping Tang, Q.-S. Gu, J.-S. Huang, Y.-P. Wang]
       {Yuping Tang$^{1}$\thanks{E-mail: tangyping@gmail.com},
        Q.-S. Gu$^{1}$\thanks{E-mail: qsgu@nju.edu.cn},
       J.-S. Huang$^{2}$,
       Y.-P. Wang$^{1}$
\\
   $^{1}$Department of Astronomy, Nanjing University, Nanjing 210093, P. R. China\\
   $^{2}$Harvard-Smithsonian Center for Astrophysics, 60 Garden Street, Cambridge, MA
   02138\\}

\begin{document}

\maketitle

\begin{abstract}

We present the Spitzer Space Telescope Infrared Array Camera (IRAC)
observations for a sample of local elliptical galaxies to study
later stages of AGN activities. A  sample of 36 elliptical galaxies
is  selected from the Palomar spectroscopic survey. We detect
nuclear non-stellar infrared emission in 9 of them. There is
unambiguous evidence of circumnuclear dust in these 9 galaxies in
their optical images. We also find a remarkable correlation between
the infrared excess emission and the nuclear radio/X-ray emission,
suggesting that infrared excess emission is tightly related with
nuclear activity. Possible origin of infrared excess emission from
hot dust heated by the central AGN is supported by spectral indices
of IR excess emission.
\end{abstract}


\section{Introduction}

The tight correlation between the velocity dispersion of the bulge
of the host galaxy and the black hole mass suggests that
supermassive black hole (SMBH) is an ubiquitous
component in elliptical galaxies (Gebhardt et al. 2000; Ferrarese \&
Merritt 2000; Kormendy \& Richstone, 1995; Magorrian et al. 1998).
Activities of SMBHs are well known for elliptical galaxies at high
redshifts, these systems are host galaxies for most
radio-loud and the bright quasars (Hutchings \& Morris 1995;
Bahcall, Kirkakos \& Schneider 1996; Falomo et al. 2005). However,
most SMBHs in nearby ellipticals are no longer very active. Ho et al.(1995,1997)
found that, in the Palomar spectroscopic survey of nearby galaxies,
only about 50\% of ellipticals show detectable
emission-line nuclei, most of which belong to be low-ionization
nuclear emission-line regions (LINERs; Heckman 1980).

Mulit-wavelength band surveys in elliptical galaxies have been
carried out to search for nuclear emission from LINERs and other
low-luminosity active galactic nuclei (LLAGNs) to study the true
ionization mechanism.  Such observations were performed in optical
(Ho, et al. 1995, 1997); X-ray (Terashima, et al. 2000, 2002; Flohic
et al. 2006), radio (Nagar et al. 2002, 2005; Filho et al. 2006),
and ultraviolet band (Maoz, et al. 1995; 1996). None of these
observation were conclusive in determining the nature of these
objects. A widely accepted perspective is that LLAGNs display
remarkably different properties with their high-luminosity
counterparts such as quasars and Seyferts. The
optical-to-ultraviolet "big blue bump", a typical Spectral Energy
Distribution (SED) of high-luminosity AGNs, is weak or absent in
LLAGNs (Ho, 1999). At the same time, an anticorrelation between
radio-loudness and Eddington ratio for AGNs is reported and
confirmed by recent studies (Ho, 2002; Greene et al. 2006; Sikora et
al. 2007). In the infrared band, the composite SEDs of LLAGNs
display a mid-infrared peak (Ho, 2008). Ho (2008) concludes that
LLAGNs have an radiatively inefficient accretion flow (RIAF) with a
truncated thin disk. Maoz (2007), however, found that SEDs for a
sample of relatively unobscured low luminosity LINERs show no
difference with that for higher luminosity AGNs. A similar
conclusion was obtained recently by Dudik et al. (2009) that
[NeV]$24\mu m$ /[OIV]$26\mu m$ mid-infrared line flux ratio for
LLANGs similar with standard AGN, arguing against a UV-to-optical
deficiency due to inefficient accretion in LLANGs.

On the other hand, recent observations show that the circumnuclear
region of elliptical galaxies is also complicated. Elliptical
galaxies are thought to contain only old stellar population and hot
gas. This picture has been challenged by recent observations.
Shields (1991) detected warm gas ($T\sim10^4K$) in many elliptical
galaxies (see also Macchetto et al. 1996). Recent neutral hydrogen
observations reveal substantial amount of neutral hydrogen gas in
many early-type galaxies (Morganti et al. 2006; Noordermeer 2006).
Moreover, even cool interstellar medium (ISM) including dust and
molecular gas were detected in elliptical galaxies (Knapp et al.
1985; Knapp et al. 1989; van Dokkum et al. 1995; Wiklind et al.1995;
Temi et al. 2004; Lauer et al. 2005; Sage et al. 2007; Kaneda et al.
2008). The circumnuclear cold interstellar media are mostly detected
in elliptical galaxies with nuclear activities (Tran et al. 2001;
Krajnovic \& Jaffe, 2002; Xilouris \& Papadakis, 2002; Sim\'{o}es
Lopes et al. 2007; Zhang et al. 2008). It is more likely that active
elliptical galaxies possess circumnuclear dust feature. This fact
challenges the RIAF accretion scenario, which assumes LLAGNs are not
gas-starving, but have a low radiative efficiency.

The Spitzer Space Telescope (Werner et al. 2004) with 2'' spatial
resolution (Fazio et al. 2004) in mid-infrared ($3.6-8.0\mu m$)
offers us a new approach to study both LLAGNs and their environment
in ellipticals. Although contributions of photospheric emission from
evolved stellar population and hot dust in the circum-stellar
envelopes of AGB stars form a considerable mid-infrared background
in ellipticals (Athey \& Bregman 2002; Xilouris et al. 2004; Temi et
al. 2008), smooth spatial distribution of surface brightness of
elliptical galaxies permit to separate non-stellar nuclear source
from galaxy component. There were several detections of mid-infrared
core feature in previous studies (Pahre et al. 2004; Gu et al.
2007). This excess emission from AGN component can not be easily
extracted from host galaxy extended component due to poor spatial
resolution.

In this paper, we present results of MIR observation of a sample of nearby
elliptical galaxies at mid-infrared wavelength  with Infrared
Array Camera (IRAC) and Multiband Imaging Photometer (MIPS) aboard
the Spitzer Space Telescope in search for
mid-infrared core, and studies of correlation between infrared core
feature and other nuclear properties of elliptical galaxies.
This MIR observation permits to
investigate the origin of nuclear infrared emission in ellipticals.

This paper is organized as follows: the sample selection and data
reduction are described in Section 2; the basic results of the
observation are presented in Section 3; the results are discussed in
Section 4; the conclusion is given in Section 5.


\section{Sample Selection and Data Reduction}


\begin{table}
\begin{center}
\caption{Global and Nuclear Properties. \label{table1}}


\begin{tabular}{cccccr}
\hline\hline
Galaxy &  D & $M_B$ &  & Nuclear  &   \\
Name  & ($Mpc$) & (mag) &Class & Dust  & Ref
  \\
(1) & (2) & (3) & (4) & (5) &(6)\\
\hline

NGC315 &    65.8 & -22.22 & L1.9 & Disk/Ring   & 3 \\
NGC410 &    70.6 & -22.01 & T & ...  & ...\\
NGC777 &    66.5 & -21.94 & S &...   & ...\\
NGC821 &    23.2 & -20.11 & A & No Dust  & 2 \\
NGC1052  &  17.8  &  -19.90 &  L1.9  & Lane/Chaotic  & 5 \\
NGC2832  &  91.6  &  -22.24  & L2: & No Dust  & 4 \\
NGC3226   & 23.4   &  -19.40 & L1.9 & Disk/Ring & 3\\
NGC3377  &  8.1   &  -18.47 & A  & Lane/Chaotic   & 1 \\
NGC3379  &  8.1   &  -19.36 & L2/T2: & Disk/Ring  & 2  \\
NGC3608  &  23.4  &  -20.16 & L2/S2: & No Dust & 2\\
NGC3610  &  29.2  &  -20.79 & A & No Dust & 2 \\
NGC3640  &  24.2  &  -20.73 & A &  No Dust  & 1 \\
NGC4125  &  24.2  &  -21.25 & T &  Disk/ring  & 1 \\
NGC4168  &  16.8  &  -19.07  &  S &  Lane/Chaotic  & 1 \\
NGC4261  &  35.1  &  -21.37  & L2 & Disk/Ring  & 3 \\
NGC4291  &  29.4  &  -20.09  & A  & No Dust & 2\\
NGC4278  &  9.7   &  -18.96  & L1.9  &  Lane/Chaotic  & 2 \\
NGC4374  &  16.8  &  -21.12  & L2 & Lane/Chaotic  & 5  \\
NGC4406  &  16.8  &  -21.39 & A  &  No Dust  & 2 \\
NGC4473  &  16.8  &  -20.10 & A  & No Dust & 2\\
NGC4552  &  16.8  &  -20.56  &  T  & Lane/Chaotic   &  2 \\
NGC4564  &  16.8  &  -19.17  &  A  & No Dust & 2\\
NGC4621  &  16.8  &  -20.60 &  A &  No Dust  & 1 \\
NGC4636  &  17.0  &  -20.72  & L1.9  &  Lane/Chaotic  & 7 \\
NGC4649  &  16.8  &  -21.43 & A &  No Dust   & 2\\
NGC4660  &  16.8  &  -19.06 & A &  No Dust   & 2 \\
NGC5077  &  40.6  &  -20.83 & L1.9 & Lane/Chaotic  & 6 \\
NGC5322  &  31.6  &  -21.46 &  L2: & Disk/Ring  & 2 \\
NGC5557  &  42.6  &  -21.17 & A &  No Dust  & 1 \\
NGC5576  &  26.4  &  -20.43 & A &  No Dust  & 2 \\
NGC5813  &  28.5  &  -20.85 & L2 & Disk/Ring  & 1 \\
NGC5831  &  28.5  &  -19.96  & A  & No Dust  & 1 \\
NGC5846  &  28.5  &   -21.36 & T  & Lane/Chaotic  & 1 \\
NGC5982  &  38.7  &  -20.89 & L2 & No Dust  & 1 \\
NGC6482  &  52.3  &  -21.75 &  T & ... & ... \\
NGC7626  &  45.6  &  -21.23 & L2 & Dust/Ring  & 3 \\
\hline
\end{tabular}

\end{center}
Notes--- Col(4):Nuclear spectral type from the Palomar survey (Ho et
al. 1997):L=LINER; S=Seyfert; T=Transition object; A=absorption-line
nuclei (inactive). Col(5):Morphology of optical circumnuclear dust,
obtained from references listed in Col(6), including three types:
nuclear dust disk or dust ring; dust lane or disorganized dust
patch; no dust. Col(6):References---(1) Tran et al. 2001; (2) Lauer
et al. 2005; (3) Gonz\'{a}lez Delgado et al. 2008; (4) Lauer et al.
2007; (5) Ravindranath et al. 2001; (6) Rest et al. 2001; (7) van
Dokkum \& Franx, 1995.
\end{table}

 A sample of 36 elliptical galaxies were selected from the Palomar optical spectroscopic survey for this study.
The Palomar optical spectroscopy survey comprises all nearby
galaxies brighter than $B_T=12.5 mag$ in the northern hemisphere (Ho
et al. 1995, 1997). This sample is statistical complete, and
contains both galaxies with and without nuclear activity while the
latter ones serve as a control sample. Multi-band observations of
nuclear region has been carried out and studied for a large fraction
of this sample, making it suitable for our study of the infrared
properties. Our sample includes 2 Seyferts, 15 LINERs, 5 transition
objects and 15 inactive galaxies.

  The IRAC Basic Calibrated Data (BCD) and MIPS Post Basic
Calibrated Data (Post BCD) of these galaxies are downloaded from the
archive of Spitzer Science Center. The IRAC BCD images were
performed with basic image processing, including dark subtraction,
detector linearization corrections, flat-field corrections, and flux
calibration. We further use the custom IDL software (Huang et al.
2004) to make the final mosaic image for each object. The absolute
flux calibration for IRAC flux densities is better than 10\% (Fazio
et al. 2004). We adopt the AB magnitude system for magnitudes and
colors throughout this paper.


  To obtain the MIR color distribution of each galaxy, we first cross-convolve each image
  by using the corresponding PSF\footnote{http://dirty.as.arizona.edu/\~kgordon/mips/conv\_psfs/conv\_psfs.html}(Gordon et al. 2008, Tom Jarrett, private communication).
For example, the color difference between image at $3.6\mu m$ and at
$8.0\mu m$ was obtained by following steps, firstly:
  \begin{equation}
  \textrm{Image(3.6')} = \textrm{Image(3.6)} \otimes \textrm{PSF(8.0)}
  \end{equation}
  \begin{equation}
  \textrm{Image(8.0')} = \textrm{Image(8.0)} \otimes \textrm{PSF(3.6)}
  \end{equation}
  Where ``$\otimes$'' means convolution, then:
  \begin{equation}
  \textrm{Color(3.6-8.0)}=(\textrm{Image3.6'} \times A_{3.6}) -(\textrm{Image8.0'} \times A_{8.0})
  \end{equation}
  Where $A_{3.6}$ and $A_{8.0}$ are PSF aperture correction factors for an infinite aperture.

  We used ellipse program in the ISOPHOT package of IRAF to perform the surface photometry of each galaxy.
  Hot pixels and foreground stars were identified by eye and masked out before isophotal fitting.

   The isophotal parameters, such as ellipticity and position angle, were measured at $3.6\mu m$ where
  the Signal-to-Noise (S/N) ratio is the highest. These parameters were then applied for the surface photometry
  at 4.5, 5.8 and $8.0\mu m$. Considering the scattered light, we employed extended sources aperture
  correction for calibration provided by Tom Jarrett\footnote{http://spider.ipac.caltech.edu/staff/jarrett/irac/calibration/}.
  Since we particularly focus on the nuclear region,
 we also obtained the nuclear flux density from a circular region within an aperture of 10'' (for NGC 1052 and NGC 3226, with an aperture of 15'')
to extract the non-stellar excess emission. The size of such a
central region is determined by the radial color distribution shown
in Figure 1.



\begin{table}
\begin{center}
\caption{Multi-wavelength Nuclear Emissions. \label{table2}}

\newsavebox{\tablebox2}
\begin{lrbox}{\tablebox}

\begin{tabular}{cccccccr}
\hline\hline
Galaxy & $L_{[OIII]}$ & $L_{15GHz}$ & $L_{2-10keV}$  & $L_{60\mu m}$ & $L_{100\mu m}$ \\
Name & ($erg/s$) & ($erg/s/Hz$) & ($erg/s$) &  ($erg/s/Hz$)& ($erg/s/Hz$) \\
(1) & (2) & (3) & (4) &(5) &(6)  \\
\hline
NGC315 & 39.38 & 30.39& 41.64  & 30.22 & 30.27\\
NGC410 & $<39.32$  & $<27.78$ & ... & 0 & 0\\
NGC777 & $<39.11$  & $<27.90$ & ... & 0 & 0 \\
NGC821 &... & ... &... & 0 & 22.45 \\
NGC1052 & 39.43 & 29.14 &40.78  & 29.53 & 29.72 \\
NGC2832 & $<39.05$ & $<28.18$ &...& ... & ...\\
NGC3226  & 38.93 & 27.55 & 39.62 & ... & ...\\
NGC3377 & ... & ... &...  & 28.04 & 28.39\\
NGC3379 & 37.73 & $<25.89$ &37.89 & 0 & 0 \\
NGC3608 & 37.80 & $<26.99$ & 38.64 & ... & ...\\
NGC3610 & ... & ... & ...  & 0 & 0\\
NGC3640 &... & ... &...&  0 & 0 \\
NGC4125 & 38.85 & $<26.85$ &38.93 & 29.70 & 30.02\\
NGC4168 &37.91 & 27.01 &... & 0 & 29.30 \\
NGC4261 & 39.70 & 29.65&40.65 & 29.07 & 29.28\\
NGC4291 & ... & ... & ... & 0 & 0\\
NGC4278 & 38.88 & 28.10 &39.96 & 28.83 & 29.27\\
NGC4374 & 39.03 & 28.79 &39.58 & 29.24 & 29.54 \\
NGC4406 & ... & ...&...& 28.57 & 29.00 \\
NGC4473 & ... & ... & ... & 0 & 0\\
NGC4552 & 38.05 & 28.30& 39.41 & 28.73 & 29.20\\
NGC4564 & ... & ... & ...  & 0 & 0\\
NGC4621 &... & ... &... & 0 & 0\\
NGC4636 & ... & ... &... & 0 & 0\\
NGC4649 & ... & ... &... & 29.43 & 29.52\\
NGC4660 & ... & ... &...& 0 & 0 \\
NGC5077 & 39.52 & ... &... & ... & ...\\
NGC5322 & 38.54 & 28.18 &... & 29.71 & 30.03\\
NGC5557 & ... & ... &...& 0 & 0\\
NGC5576 & ... & ... & ...  & 28.88 & 29.20\\
NGC5813 & 38.35 & 27.37 &... & 0 & 0\\
NGC5831 &... &...&...  & ... & ...\\
NGC5846 & 38.32 & 27.79&39.54 & 0 & 0\\
NGC5982 & 38.55 & ... &...& 0 & 29.77\\
NGC6482  & ... & $<27.52$ &39.40 & 0 & 0\\
NGC7626 & 38.52 & 29.00 &... & 0 & 0\\
\hline
\end{tabular}

\end{lrbox}

\scalebox{0.85}{\usebox{\tablebox}}

\end{center}
Notes---Col(2): Luminosity of $[OIII]\lambda5007$ taken from Ho et
al. (1997), except NGC 1052, NGC 4125 and NGC 5813, which are based
on observations under non-photometric conditions. Ho et al. (2003b)
give their updated $H\alpha$ luminosities and we use them
 to derive their $[OIII]\lambda5007$ luminosities, assuming same ratios of
 $[OIII]\lambda5007/H\alpha$ from Ho et al.(1997). Col(3): Nuclear Luminosity
 at 15GHz, taken from Nagar et al.(2005), except NGC1052, which is
 taken from Kellermann et al.(1998). Col(4): Nuclear Luminosity at 2-10keV, taken from Gonzalez-Martin et al. (2006), except
 NGC 4278, which is taken from Terashima \& Wilson (2003).  Col(5) - (6): Total
 luminosities at $60\mu m$ and $100\mu m$ from IRAS observation (Knapp et al. 1989)
\end{table}

\section{Results}

Figure 1 shows color distributions of $[3.6]-[4.5]$, $[3.6]-[5.8]$,
$[3.6]-[8.0]$ for all elliptical galaxies in our sample. In $10''
<R<40''$, the infrared colors for almost all galaxies do not change
significantly,  and are generally consistent with photospheric
emission of late-type stars with a minor contribution of hot dust in
circum-stellar envelopes of AGB stars (Pahre et al. 2004; del Burgo,
Carter \& Sikkema, 2008; Temi, Brenghti \& Mathews, 2008). However,
the colors in the central region with $R<10''$ tell a different
story: 9 out of 36 galaxies exhibit much redder colors. The
[3.6]$-$[8.0] shows most color excess in the central region. Thus we
use [3.6]$-$[8.0] colors to distinguish galaxies with significant
excess emission from normal galaxies. For galaxies with no
significant color excess, the deviation of [3.6]$-$[8.0] color in
central 10'' region is only $-1.44 \pm 0.02$. The remaining 9
galaxies all have [3.6] - [8.0] color redder than -1.34, indicating
a redder central color above $3\sigma$ level.

\begin{figure}
\includegraphics[width=9cm]{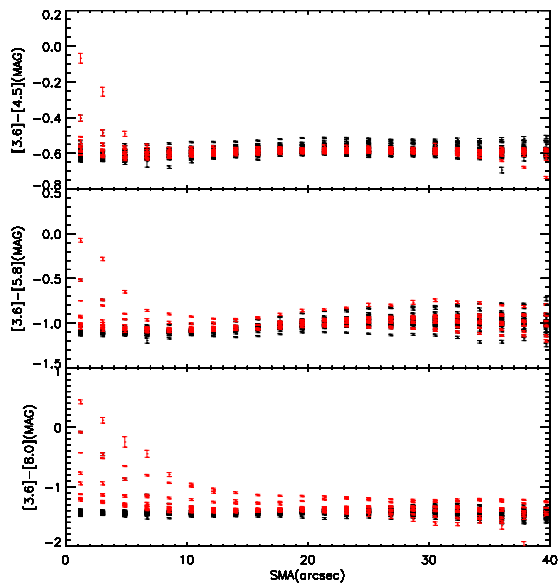}
\caption{Color distributions of all non-core galaxies (black) in
comparison with core (red) galaxies, the x-axis refers to distances
from the center along major axis, in units of arcsec. Core galaxies
have obvious redder color in central several arcseconds, especially
seen from $3.6 - 8.0\mu m$.} \label{fig01}
\end{figure}

Figure 2 is the color-color diagram for galaxies with no color
excess in the center. The dispersions of their IRAC colors are very
small, with $[3.6]-[4.5]=0.61\pm0.02$, $[3.6]-[5.8]=-1.11\pm0.02$,
$[3.6]-[8.0]=-1.44\pm0.02$, in agreement with colors of M-type star
( Pahre et al. 2004).  This result supports that normal ellipticals
are dominated by old stellar population. Moreover, LLAGNs and
inactive galaxies show no systematic difference in mid-infrared
color, which agrees with the nuclear stellar population analysis of
LLAGNs in previous studies (Boisson et al. 2000; Ho et al. 2003b).
Zhang et al.(2008) studied the nuclear stellar population for a
sample of early-type galaxies that is highly overlapped with our
sample, and they found no difference in stellar age distribution
between LLAGNs and inactive galaxies in their sample.


The mid-infrared emission from elliptical galaxies consists of two
components: stellar and non-stellar emission (dust, AGN, etc.). The
$3.6\mu m$ emission is dominated by later-type stellar photospheric
emission (Pahre et al. 2004; Temi et al. 2008), thus traces stellar
mass distribution. The non-stellar component becomes significant at
longer wavelength, which is shown clearly by the $[3.6]-[8.0]$
color. To derive the flux density of excess non-stellar emission in
infrared core galaxies, we use the mean colors ([3.6]-[5.8],
[3.6]-[8.0]) for central regions of non-core galaxies as a template
of the old stellar component. For five infrared core galaxies with
visible excess emissions (NGC 315, NGC 1052, NGC 3226, NGC 4261 and
NGC 5322) at both $5.8\mu m$ and $8.0\mu m$, by assuming that the
shape of SED for the central excess emission as a power-law
function, we are able to disentangle excess component from stellar
emission by solving following equations:
  \begin{equation}
  f_{exs,i}+f_{star,i}=f_{tot,i}
  \end{equation}
  \begin{equation}
  f_{star,i}/f_{star,j}=R_{i,j}
  \end{equation}
  \begin{equation}
  f_{exs,i}/f_{exs,j}=(\nu_i/\nu_j)^\alpha
  \end{equation}
  where subscript numbers $i/j=1,2,3$ correspond to different wavelength
   $\lambda_{1,2,3}=3.6, 5.8, 8.0$, $\nu_{i/j}$ are corresponding frequency of $\lambda_{i/j}$,
  $f_{exs,i/j}$ and $f_{star,i/j}$ are flux densities
  of core emission and stellar emission at different bands,
  respectively. $f_{exs,i/j}$, $f_{star,i/j}$ and $\alpha$ are set as variables,
with a total number of 7, equal to the number of equations.
$R_{i,j}$ are flux ratios of emission for old stellar population,
obtained by averaging central colors of all non-core galaxies.
$f_{tot,i}$ are flux densities of stellar emission at different
bands, extracted from central region within an aperture of 15'' for
NGC 1052/ NGC 3226 and 10'' for other core galaxies. The size of
aperture is determined by two factors, it should be large enough to
include the whole infrared core structure, and as small as possible
to reduce the influence from offset of zero-point caused by
uncertainty in true nuclear stellar color, this influence bring
larger uncertainty at shorter wavelength, where dilution by stellar
emission is more serious. The special resolution of IRAC is 2''. In
[3.6 - 8.0] distribution shown in Figure 1, except NGC 1052, NGC
3226 and NGC 4278, the sizes of red core structures for all the rest
six galaxies are smaller or close to 10 arcsec. Furthermore, due to
the convolution procedure mentioned in Section 2, the sizes of core
structure in Figure 1 appear larger than their real values.
Therefore, we consider 10 arcsec a proper size for the aperture to
extract nuclear excess emission. For NGC 1052 and NGC 3226, 15''
aperture is adopted, NGC4278 show a redder color throughout the
galaxy and a vague core structure, we simply use 10'' aperture to
derive a "nuclear" flux in this object. For such apertures, the
propagated relative errors of excess emission due to uncertainties
of $R_{ij}$ are several to about ten percent.

\begin{figure}
\includegraphics[width=9cm]{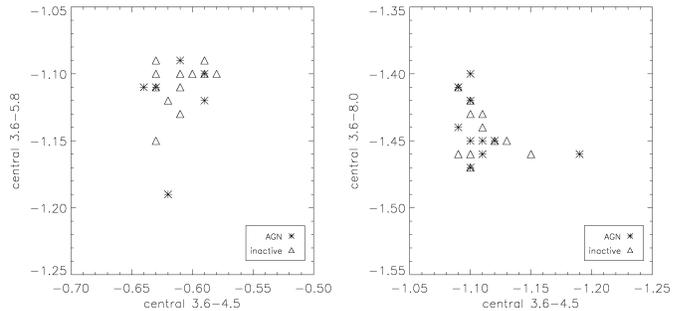}
\caption{Central colors of non-core galaxies within 10'', stars are
LLAGNs while triangles are inactive galaxies.} \label{fig02}
\end{figure}

Flux densities and luminosities obtained through above approachs are
listed in Table 3 and 4. In Table 3, the errors of total IRAC and
MIPS flux densities of whole galaxies are simply set as $10\%$
(Fazio et al. 2004; Rieke et al. 2004), which is an estimation for
the uncertainties of absolute calibration, in contrast,
uncertainties derived from image statistics are neglibile. The
uncerntainties of excess emission and spectral indices are estimated
from deviation of excess emissions extracted from $\pm 5''$
apertures (10''/15''/20'' for NGC 1052 and NGC 3226, 5''/10''/15''
for other galaxies). The proportion of excess emission at 3.6 is
very small, contributes less than 5\% even in the strong infrared
core galaxy, NGC 1052, and less than 0.5\% in NGC 4261. Hence it is
reasonable to assume all $3.6\mu m$ flux densities in the other four
galaxies (NGC 4125, NGC 4278, NGC 4374 and NGC 5077) are produced by
stellar component, with this assumption we obtain the excess flux
densities at $8.0\mu m$ for these four galaxies, since their excess
emissions at short wavelength, even if exist, will be too weak to be
detected. The excess emissions are generally small comparing with
stellar emission except in NGC 1052, where $8.0\mu m$ excess
emission contributes to nearly 50 percents of total emission.

\begin{figure*}
\includegraphics[width=12cm]{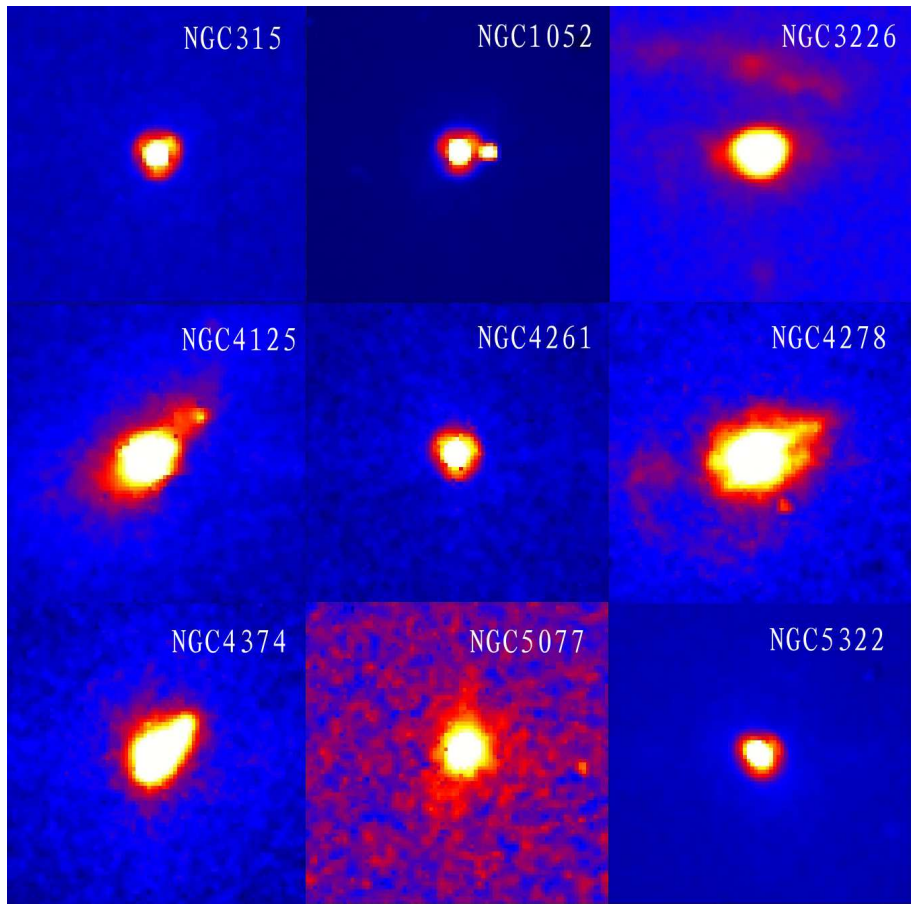}
\caption{Residual images of infrared core galaxies at $8\mu m$ in
central $24'', \times 24''$ regions, with the background from old
stellar population removed. Point-like structure can be seen in four
galaxies: NGC315, NGC1052, NGC4261 and NGC5322, while the rest
galaxies show extended excess emission at $8\mu m$. The substructure
in NGC 1052 is a bandwidth effect.} \label{fig03}
\end{figure*}

Figure 3 shows $8.0\mu m$ residual images of nine infrared core
galaxies, which are obtained by utilizing $3.6\mu m$ image and flux
ratio $R_{3.6, 8.0}$ appearing in equation (5) to remove the
contribution of underlying stellar population. The $8\mu m$ excess
emissions in four galaxies (NGC 315, NGC 1052, NGC 4261 and NGC
5322) show point-like structure with ring-like feature of $8.0\mu m$
Point Spread Function (PSF), and the other five galaxies show
extended excess emission, indicating off-nuclear sources of $8\mu m$
excess. NGC 1052 show a substructure on the right of the center.
This is due to the bandwidth effect\footnote{IRAC Data Handbook:
http://ssc.spitzer.caltech.edu/irac/dh/} and was masked for
measuring photometry.

To show the contribution of extended emission, in Figure 4 we
compare nuclear $8\mu m$ excess luminosity with the compactness of
excess emission, quantified by  $L(<5'')/Lex,total$, the proportion
of nuclear excess luminosities within 5'' aperture to total extended
luminosities. For point-like sources, the surface brightness profile
within 5'' are uniform and consistent with $8\mu m$ PSF. The
proportion of central emission within 5'' are generally higher than
$50\%$ and are higher than $80\%$ for point-like sources. The
compactness decreases as nuclear luminosity decreases. Extended
emissions become considerable only for sources with nuclear
luminosities lower than $3 \times 10^{27} erg/s/Hz$.

 \section{Discussion}
  The excess non-stellar emission may originate from the central
  AGNs or nuclear hot dust heated by AGN.
 PAHs emission feature could also contribute to excess emission at
$8.0 \ \mu m$, this is exactly the case for NGC 1052, where $7.7\mu
m$ PAHs emission feature has been detected (Kaneda et al. 2008),
though this feature by itself is not able to explain the excess
emission at $4.5\mu m$ and $5.8\mu m$. AGNs and kpc-scale
circumnuclear dust are commonly detected in nearby elliptical
galaxies. In the Palomar spectroscopic survey of nearby galaxies,
about half of the ellipticals show detectable emission-line nuclei,
while most of which are classified as LINERs (Ho, et al. 1997). On
the other hand, recent observations from Hubble Space Telescope
(HST) images show that circumnuclear dust appear in about $\sim40\%$
of ellipticals in their optical images (Tran, et al. 2001; Lauer, et
al. 2005; Simoes Lopes, et al. 2007). PAHs were thought to be rare
in ellipticals considering the sputtering destruction in the hot
plasma environment. However, recently, Kaneda et al.(2008) reported
detection of PAHs emission features in 14 out of 18 dusty
ellipticals, implying that PAHs are more common than ever thought in
these systems. In the following discussion, we confine our interest
to the role of different components playing in produce the excess
infrared emission.

  Columns (4) and (5) of Table 1 summarize the classification on nuclear activity and circumnuclear
  dust morphology in optical images of our sample galaxies. All infrared core galaxies,
  both point-like and extended sources, show AGN activities, 8 of them are LINERs, except for NGC 4125,
  which is classified as a Transition object. Thus infrared core galaxies account for
  $41\%$ of AGNs in our sample. With respect to circumnuclear dust, 17 galaxies, about half of our sample,
  have been detected of circumnuclear dust in optical band, all infrared core galaxies belong to this group.
  Therefore, the fact that both AGNs and circumnuclear dust coincide with a central infrared core makes it
  difficult to clarify their contributions to the infrared excess emission.

\begin{table*}
\begin{center}
\caption{Total and Excess Infrared Flux Densities.\label{table3}}

\newsavebox{\tablebox3}
\begin{lrbox}{\tablebox}

\begin{tabular}{lccccccccccc}

\hline\hline
Galaxy & $S_{T3.6}$ & $S_{T4.5}$ & $S_{T5.8}$ & $S_{T8.0}$ & $S_{T24}$ & $S_{E3.6}$& $S_{E5.8} $ & $S_{E8.0}$  \\
Name   & (mJy)      & (mJy)      & (mJy)      & (mJy)      & (mJy)     & (mJy)     & (mJy)       & (mJy) & $\alpha_{3.6-8}$ & $\alpha_{8-24}$ & $\alpha_{radio}$ \\
(1)   &(  2)        &(3)         &(4)         &(5)         &(6)        &(7)        &(8)          &(9)    & (10)             & (11)            & (12) \\
\hline NGC 315 & $207.7\pm20.7$ & $122.7\pm12.2$ & $74.0\pm7.4$ &
$51.4\pm5.1$ & $100.1\pm10.0$ & $1.3\pm0.3$ & $4.4\pm0.5$ &
$10.0\pm0.6$ & $2.5\pm0.1$& 2.1 & 0.27 \\
NGC1052 & $337.7\pm33.7$ & $207.0\pm20.7$ & $163.4\pm16.3$ &
$136.1\pm13.6$ & $...$ & $17.6\pm0.5$ & $38.7\pm0.5$ & $65.9\pm0.4$
& $1.8\pm0.1$ & 1.3$^a$ & 0.33 \\
NGC4261 & $355.6\pm35.5$ & $209.1\pm20.9$ & $134.6\pm1.3$ &
$83.8\pm8.3$ &$50.5\pm5.0$& $0.8\pm0.1$ & $2.0\pm0.2$ & $3.9\pm0.3$
& $2.0\pm0.1$ & 2.3 & -0.24\\
NGC5322 & $360.2\pm36.0$ & $216.2\pm21.6$ & $146.6\pm14.6$ &
$90.6\pm9.0$ & $33.5\pm3.3$ & $1.3\pm0.4$ & $4.0\pm0.6$ &$8.4\pm0.4$ & $2.3\pm0.5$ & 1.2 & 0.15\\
NGC3226 & $117.1\pm11.7$ & $61.7\pm6.1$ & $44.1\pm4.4$ &
$33.5\pm3.3$ &$30.1\pm3.0$ & $1.4\pm0.2$ & $4.6\pm0.3$ &
$10.2\pm0.4$ & $2.3\pm0.2$ & 1.0 & 0.11\\
NGC4125 & $541.2\pm54.1$ & $309.2\pm30.9$ & $211.0\pm21.0$ &
$122.5\pm12.2$ &  $31.9\pm3.2 $ & ... & ... & $4.5\pm0.8$ & ... & 1.8 & ...\\
NGC4278 & $386.3\pm38.6$ & $233.2\pm23.3$ & $149.7\pm14.9$ &
$110.6\pm 11.0$ & $29.1\pm2.9$ & ... & ... & $11.0\pm2.0$ & ... & 0.9 & 0.29\\
NGC4374 & $773.6\pm77.3$ & $462.9\pm46.3$ & $284.9\pm28.4$ &
$180.4\pm 18.0$ & $28.9\pm2.8$ & ... & ... & $6.7\pm1.3$ & ... & 1.3 & 0.13\\
NGC5077 & $162.6\pm16.3$ & $94.1\pm9.4$ & $79.9\pm8.0$ &
$37.8\pm 3.7$ & $21.5\pm2.1$ & ... & ... & $1.9\pm0.5$ & ... & 2.2 & ...\\

\hline

\end{tabular}

\end{lrbox}

\scalebox{0.85}{\usebox{\tablebox}}

\end{center}

Notes---Col(2)---(6):Total infrared flux density at
$3.6$,$4.5$,$5.8$,$8.0$ and $24\mu m$. Col(7)---(9):Flux density of
central excess emission at $3.6$, $5.8$, $8.0\mu m$. Col(10):
Power-law index of central excess emission between $3.6-8\mu m$,
derived from $3.6$, $5.8$ and $8.0\mu m$ through method described in
Section 3. Col(11): Power-law index of central excess emission
between $8-24\mu m$, derived from Col(6) and Col(9). Col(12):
Spectral index of nuclei derived from flux densities at 5GHz and
15GHz, which are taken from Nagar et al. (2005).

$^a$ Since NGC1052 has no MIPS data, the power-law index is derived
from the mid-infrared spectra, that is given by Kaneda et al, 2008.

\end{table*}


\begin{table*}
\begin{center}
\caption{Total and Excess Infrared Luminosities.\label{table3}}

\newsavebox{\tablebox4}
\begin{lrbox}{\tablebox}

\begin{tabular}{lcccccccc}

\hline\hline Galaxy & $L_{\nu,T3.6}$ & $L_{\nu,T4.5}$ &
$L_{\nu,T5.8}$ & $L_{\nu,T8.0}$ & $L_{\nu,T24.0}$ & $L_{\nu,E3.6}$&
$L_{\nu,E5.8} $ &
$L_{\nu,E8.0}$ \\
Name & ($10^{28}erg/sec$) & ($10^{28}erg/sec$) & ($10^{28}erg/sec$) & ($10^{28}erg/sec$) & ($10^{28}erg/sec$) & ($10^{28}erg/sec$) & ($10^{28}erg/sec$) & ($10^{28}erg/sec$)  \\
(1) &(2)&(3)&(4)&(5)&(6)&(7)&(8)& (9)  \\
 \hline
NGC315 & $107.61$ & $63.57$ & $38.35$  & $26.64$ & $51.28$ & $0.68$
& $2.26$ & $5.08$
\\
NGC1052 & $12.80$ & $7.85$ & $6.19$ & $5.16$ & ... & $0.67$ & $1.47$
& $2.50$
 \\
NGC4261 & $52.41$ & $30.82$ & $19.84$ &
$12.35$ & $7.41$ & $0.11$ & $0.35$ & $0.57$   \\
NGC5322 & $43.03$ & $25.83$ & $17.52$ &
$10.82$ &$3.98$ & $0.16$ & $0.47$ &$1.00$  \\
NGC3226 & $7.67$ & $4.04$ & $2.89$ & $2.20$ & $1.95$ & $0.09$
& $0.30$ & $0.67$ \\
NGC4125 & $37.92$ & $21.67$ & $14.78$ &
$8.58$ & $2.24$ & ... & ... & $0.31$ \\
NGC4278 & $4.35$ & $2.63$ & $1.69$ &
$1.24$ &$0.33$ & ... & ... & $0.12$ \\
NGC4374 & $26.12$ & $15.63$ & $9.62$ &
$6.09$ & $0.97$ & ... & ... & $0.22$  \\
NGC5077 & $32.07$ & $18.55$ & $15.71$ &
$7.45$ &$4.27$ & ... & ... & $0.38$  \\

\hline
\end{tabular}

\end{lrbox}

\scalebox{0.9}{\usebox{\tablebox}}

\end{center}

Notes---Col(2)---(5):Total infrared luminosity at $3.6$, $4.5$,
$5.8$, \& $8.0\mu m$. Col(6)---(8):Luminosity of central excess
emission at $3.6$, $5.8$, $8.0\mu m$.

\end{table*}

  Optical observations of early-type galaxies show that active early-type galaxies more tend to possess circumnuclear dust
  than inactive ones (Tran et al. 2001; Krajnovic \& Jaffe, 2002;
  Xilouris \& Papadakis, 2002; Simoes Lopes et al. 2007).
  Simoes Lopes et al. (2007) reported that 100\% AGNs show ciucumnuclear dust feature in optical image, in contrast, only
  2 out of 15 ellipticals are detected with dust. This correlation can be understood in two ways. Firstly,
  the accretion of a black hole requires fuel, and circumnuclear dust is a good indication of cold gas
  inflow and thus an evidence of fuel supply for the central of AGN. Nevertheless,
  the existence of ellipticals without visible dust but hosting AGNs, although in small amount: NGC 2832, NGC 3608, NGC 5982,
  exclude the presumption that nuclear dust is necessary for nuclear activity.
  On the other hand, Temi et al. (2007a, b) suggested that AGN feedback may play an important role in
  transporting dust from central reservoir to the interstellar space in ellipticals,
  this mechanism may also assist to establish such a correlation.

    Figure 5 shows plots of $60\mu m$ luminosity versus $100\mu m$ luminosity based on IRAS data
  (Knapp et al. 1989),
  most of the far-infrared detected galaxies are infrared core galaxies hosting luminous AGN, while 4 inactive
  galaxies and 1 non-core AGN also show detection but lower luminosities on the
  whole. This is consistent with the fact that AGN nuclei preferentially exist in
  optically dusty ellipticals.
  Although for the most powerful AGNs: NGC 1052, NGC 315, NGC 4261, the far-infrared emission may come
  from strong radio continuum powered by AGN, it is unlikely that this is the dominant contribution
  for other fainter AGNs, since their far-infrared emission are too strong comparing with the
  extrapolation of radio continuum. Additional evidence arises from resolved far-infrared
  emission in NGC 5077 (Temi et al. 2007b) and equal dust mass between that inferred from far-infrared measurement and
  that from optical extinction observation
  in NGC 4125 (Bregman et al. 1998). In dusty ellipticals the dust mass inferred from far-infrared
  is generally an order of magnitude greater than that from optical
  extinction observation (Goudfrooij et al. 1994), indicating that the distribution of dust is extended.
  Therefore, the higher far-infrared luminosities in luminous AGNs than faint AGN and inactive galaxies reveal that correlation between
  AGN and dust does not only hold for dust at circumnuclear region but also for extended dust throughout the galaxy.

\begin{figure}
\includegraphics[width=9cm]{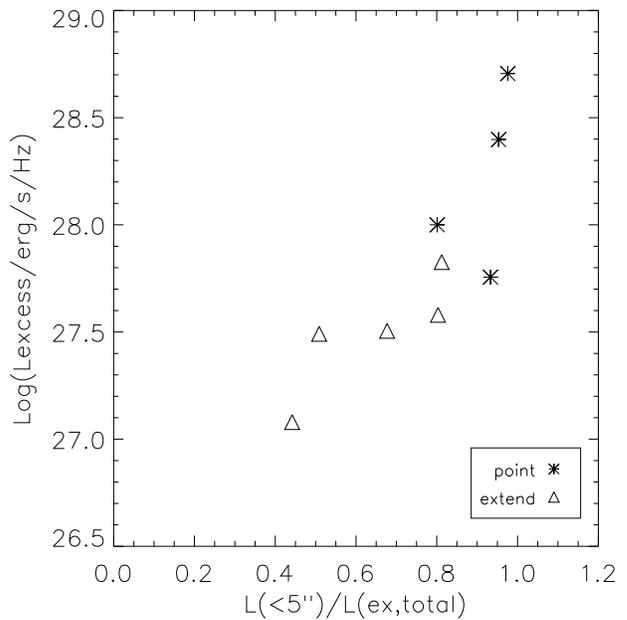}
\caption{Nuclear excess luminosity plotted against the proportion of
nuclear excess luminosity  within 5'' aperture to total extended
luminosity} \label{fig04}
\end{figure}


\begin{figure}
\includegraphics[width=9cm]{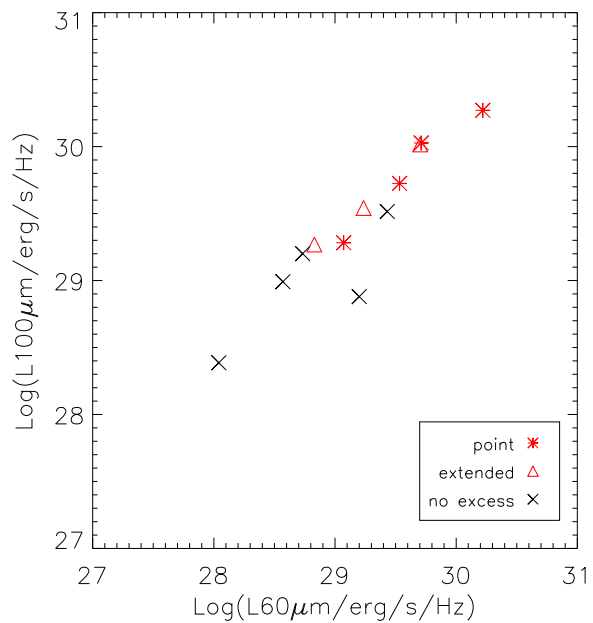}
\caption{Far infrared luminosities at $60\mu m$ and $100\mu m$,
symbols have same meaning as figure 4. Stars denote point-like core
galaxies, triangles denote extended core galaxies and crosses denote
other LLAGNs without infrared core. } \label{fig06}
\end{figure}


\begin{figure}
\includegraphics[width=9cm]{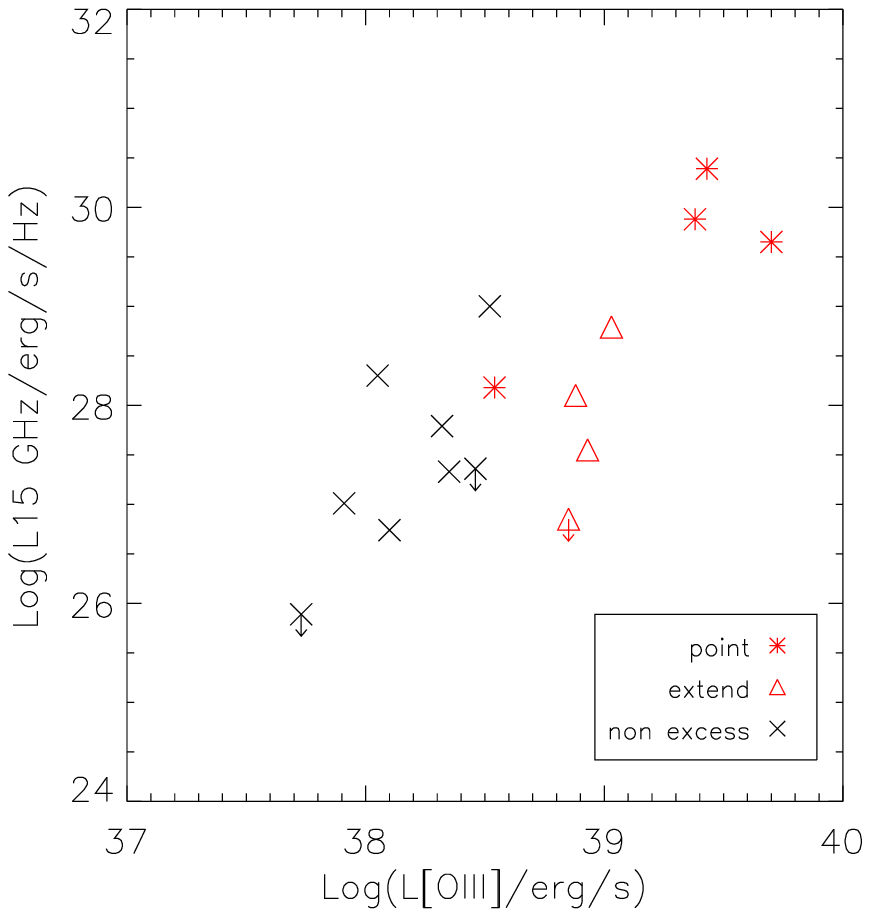}
\caption{Nuclear $[OIII]\lambda5007$ luminosity plotted against
15GHz radio luminosity, symbols with red color refer to infrared
core galaxies, stars are point-like core galaxies, triangles are
extended core galaxies, and crosses are other LLAGNs without
infrared core. } \label{fig05}
\end{figure}

  Figure 6 shows the 15 GHz nuclear radio luminosity versus nuclear $[OIII]\lambda5007$
  luminosity for all AGNs in our sample. The nuclear 15 GHz radio luminosities
are mainly taken from Nagar et al. (2005),
  and the nuclear $[OIII]\lambda5007$ luminosities are taken from Ho et al.(1997), which have been corrected for reddening effect by using the Balmer decrement.
  It is obvious that infrared core AGNs distinctly separate from non-core AGNs with higher luminosities both at optical and radio
  wavelength.
  Especially, for $[OIII]\lambda5007$ emission, core AGNs and non-core AGNs are completely distinguished at
  $\log(L_{O[III]})\sim  38.5$.
  With respect to radio emission, 7 out of 9 core galaxies have been detected of compact radio nuclei,
  all of which exhibit flat spectrum---as shown in the last column of Table 3---and high brightness temperature with
  $T>10^7K$ indicating non-stellar origin.

  Figure 6 strongly suggests that the activity of central black hole accounts for infrared excess emission.
  If infrared excess luminosity is correlated with AGN luminosity,
  a infrared-red core will intrinsically exist for every AGN, but be too weak to be detected in cases of low
  luminosity.
  The four panels of Figure 7 show $8.0\mu m$ excess
  luminosity plotted against nuclear 15GHz radio luminosity, MIPS $24\mu m$
  luminosity, nuclear $[OIII]\lambda5007$ luminosity and nuclear X-ray luminosity at
  2-10keV.
  The linear fit of data points yields linear-correlation
  R values of $0.65$,  $0.92$, $0.32$ and $0.78$, respectively.
  While the correlation between excess $8\mu m$ emission and $24\mu
  m$ emission can be simply attributed to similar origins, correlations
  between $8\mu m$ excess and radio and hard X-ray emission,
  further support an AGN origin for infrared core.
  In comparison, the weak correlation between excess emission and $[OIII]\lambda5007$
  luminosity might be resulted from both extinction and shock ionization.
   A possible interpretation for the considerable scatter in Figure 7 is unresolved non-AGN sources.
   Infrared core galaxies with weak excess emission
  in Figure 3 generally have extended structure, while higher-luminosity sources mainly display point-like morphology, suggesting non-AGN contribution
  in low luminosity sources. Particularly, the
  weak emission at $8\mu m$ in extended sources imply PAHs
  origin, which are generally detected in dusty ellipticals (Bregman et al. 2006;
  Kaneda et al. 2008). Yet, PAHs emission, even if exist (For example: NGC 1052, Kaneda et al. 2008), will still play a
  minor role in strong and point-like sources considering the
  remarkable emission at wavelength shorter than $8\mu m$ in these
  objects.

\begin{figure*}
\includegraphics[width=18cm]{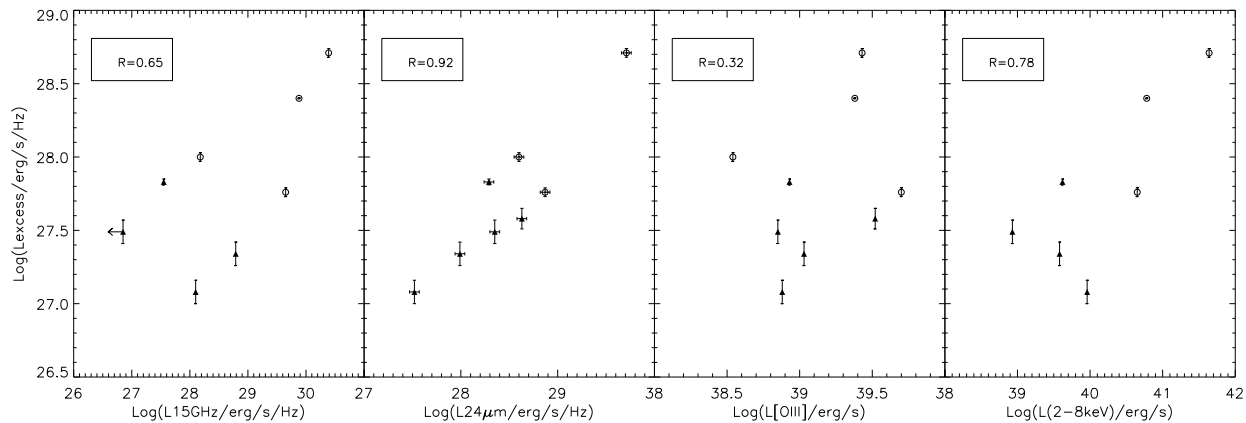}
\caption{Comparison between $8\mu m$ excess emission and a set of
AGN indicators. Open circles denote point-like sources and filled
triangles denote extended sources. Y-axis refers to luminosities of
excess emission at $8\mu m$ and x-axis refers to 15GHz core
luminosities, MIPS $24\mu m$ luminosities, nuclear
$[OIII]\lambda5007$ luminosities, and nuclear luminosities at
2-10keV, from left to right. The linear-correlation coefficient R is
also shown.} \label{fig07}
\end{figure*}

  The contribution of PAHs can be further evaluated by
  comparing spectral indices at long and short wavelength. Since
  uncertainties of excess emission are serious at wavelength shorter than $5.8\mu
  m$, here we only consider three galaxies (NGC 315, NGC 1052, NGC 3226) given their reliable measurement at short wavelength. The approach
  to derive excess flux density described in Section
  3 is based on the assumption that excess emission through IRAC band
  satisfy a single power-law distribution with a uniform $alpha$. In order to compare
  spectral indices between different IRAC bands, the flux density need
  to be reevaluated in another way.
  For NGC 315 and NGC 3226, the proportion of non-stellar component at $3.6\mu m$
  flux density is negligible, thus we assume that all $3.6\mu m$
  emission is due to old stellar population, following this
  assumption the derived spectral index $\alpha_{4.5-5.8}$ is 2.0
  for NGC 315 and 3.4 for NGC 3226, while $\alpha_{5.8-8.0}=2.8$ for
  both galaxies. The spectral indices are similar enough to rule
  out the possibility of dominant PAHs emission. For NGC 1052,
  we simply derive the spectral index $\alpha_{3.6,4.5,5.8}$
  following the same approach described in Section 3, the derived
  value is 1.7, also consistent with $\alpha_{3.6,5.8,8.0}=1.8$.

    In normal elliptical galaxies, mid-infrared emission is
  dominated by circumstellar hot dust around AGB stars (Knapp et al. 1992; Temi et al.
  2007a). Nevertheless, the tight correlation between MIPS $24\mu m$ emission and $8\mu
  m$ emission in Figure 7 suggests that the sources of both are the same in infrared core galaxies. To examine
  the origin of MIPS $24\mu m$ emission, we compare $24\mu m$
  luminosities with optical B-band absolute magnitude $M_B$ in Figure 8. As Temi
  et al. (2007a) noted in their work, for non-active ellipticals,
  $M_B$ scales with $24\mu m$ luminosity, supporting a stellar origin for $24\mu m$ emissions.
  Half of galaxies with $8\mu m$ excess emission also show $24\mu m$ excess with respect to normal galaxies,
  while the other four galaxies do not show obvious signs of $24\mu m$
  excess. Thus the tightness of the correlation between MIPS $24\mu m$ emission and $8\mu
  m$ emission should be taken with care when it comes to faint
  sources, which could be diluted by circumstellar emission.

  Three AGN-associated mechanisms can be responsible for the observerd excess infrared emission.
  The first is thermal emission from nuclear hot dust heated by
  central AGN, which could be expected in normal Seyfert galaxies. As discussed earlier, in active
  ellipticals, optically observed circumnuclear dust feature is common , generally with scales of a few hundreds pc.
  With such distance from nuclei, dust could not be heated sufficiently by AGN to generate observed emission at IRAC
  band (van Bemmel et al. 2003 \& 2004). Hot dust responsible for observed
  emission, if exist, will lie within a central region of several to about ten pc,
  where is traditionally assumed as torus and is difficult to be resovled. However, the viewpoint that LLAGNs possess torus as Serferts
  is challenged by the observed low X-ray column densities in LLAGNs (Terashima et al. 2002;
Gonz¨¢lez-Mart¨ªn et al. 2006) and high detection rate of optical
compact core (Chiaberge et al.
  1999), which indicates mild obscuration. The disappearance of dust emission is
  also predicted by "wind-torus" model, in which obscuring torus is essentially clumpy dusty wind
emanating from accretion disk and could not maintain while accretion
rate declines to the level insufficient for out flow (Emmering et
al. 1992; Konigl \& Kartje, 1994; Elitzur et al. 2006). Without hot
dust, a second choice to interpret the infrared excess is the
synchrotron emission from thermal electrons
  in accretion flows, existing models of RIAF predict a submillimeter to infrared bump in the
  SED (Quataert et al. 1999; Yuan et al. 2003). Finally, it is worthwhile to notice that 8 out of 9 core galaxies in our sample present
  compact radio core, and 4 of which (NGC 315, NGC 1052, NGC 4261, NGC 4374) belong to the category of FR I radio galaxies with
  kpc scale jets. There is a long-time consideration that the SED of FR I radio
  galaxies, or even other types of LLAGNs, could be dominated by synchrotron emission from
  the jet (Chiaberge et al. 1999; Yuan et al. 2002; Falcke et al.
  2004). One typical example of jet dominated LLAGN is M87 (Shi et al. 2007; Perlman et al.
  2007), where emission from bright knots of jet could be seen at all IRAC bands and nuclear mid-infrared emission is primarily due to the
  non-thermal emission from the base of the jet. Yet, unlike M87, none of core
  galaxies in our sample
  show any signs of jet emission at IRAC band, and our core
  galaxies generally have radio luminosities lower than M87 by a factor of 100, but comparable
  infrared excess emission.

\begin{figure}
\includegraphics[width=7cm]{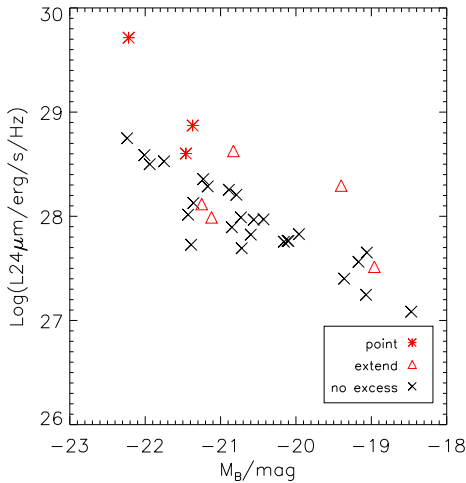}
\caption{MIPS $24\mu m$ luminosity plotted against optical B-band
absolute magnitude. Symbols have the same meanings as in Figure 5
and Figure 6.} \label{fig08}
\end{figure}

  As shown in Table 3, for 5 sources in our sample with
  detectable excess emission throughout the IRAC band, we give their
  spectral indices $\alpha_{3.6-8}$ following the method described in Section 3. The
  spectral indices of the 5 galaxies are similar, with values
  $\sim$ 2, implying a common origin of infrared excess. Such an index
  value is too steep for a jet dominated SED, which is not expected
  to be significantly larger than 1.0 (Markoff et al. 2003; Perlman et al. 2007),
  and too flat for the model of radiation from RIAF, which lies around 3 (Yuan et al. 2003 and private communication with
  Yuan).
  A combination of this two components might interpret our derived
  indices. which could be expected in FR I radio galaxies (Wu et al.
  2007). However, it should be noticed that the emission characters
  at radio wavelength of this 5 core galaxies are not uniform, while
  3 galaxies (NGC 315, NGC 1052, NGC 4261) are FR I galaxies with large scale jets and strong radio emission,
  NGC 3226 and NGC 5322 only show relatively faint compact core and show no extended
  radio emission (Nagar et al. 2005). Thus it is questionable why the galaxies with
  distinctive radio-loudness could produce similar spectral energy
  distribution through hybrid emission including jet and accretion
  flows.
  On the contrary, this value is consistent with that of general spectral index
  of Seyfert nuclei (Alonso-Herrero et al. 2003), supporting
  origin of thermal emission from hot dust. In addition, Barth et al. (1999) detected polarized broad
  $H_\alpha$ emission in and only in the 3 FR I galaxies mentioned above (NGC 315, NGC 1052, NGC 4261),
  out of 14 LLAGNs, and all of them also show considerable
  absorption X-ray columns ($\sim 10^{22} cm^{-2}$) (NGC 1052: Guainazzi et al. 2000; NGC 4261: Sambruna et al.
  2003; NGC 315: Gonz¨¢lez-Mart¨ªn et al. 2006 ), indicating the
  existence of obscuring structure and at least some contribution of
  thermal emission. Since these three objects have relatively higher
  AGN luminosities, it is possible they more closely resemble classical AGNs than other objects
  in our sample. Therefore, while we are more inclined to attribute the infrared excess of
  relatively higher luminosity AGNs to dust emission, for evidence of torus in some of them,
  and for their similar spectral indices with Seyfert2s, the infrared emission mechanism of
  fainter objects is more uncertain. Our results show that infrared excess
  emission decreases with the decrease of other AGN indicators, and current data show no sign for a change of infrared
  emission mechanism.
  On the other hand, although LLAGNs generally have low X-ray column density,
  X-ray absorption does not have a direct connection with infrared emission.
Firstly, the X-ray inferred column density is determined by
obscuring material along our line of sight, while the infrared
emission is integrated emission emission from all sources
surrounding AGN. Secondly, while both produce X-ray absorption, the
contribution of dust-free gas is larger as compared to dusty gas.
The important role of dust-free gas in X-ray absorption is supported
by lower column density inferred from reddening effect than that
from X-ray absorption. (Maccacaro et al. 1982; Maiolino et al.
2001). The insignificant X-ray absorption does not automatically
indicate disappearance of dust emission. By assuming Galactic
dust-to-gas ratio and "standard" model of interstellar dust, the
column density $N_H \sim 2 \times 10^{21} \tau_V cm^{-2}$ (Elitzur
2008), a neutral hydrogen column density with a few $10^{21}
cm^{-2}$ could still cause a considerable absorption in optical/UV
band and thus thermal re-emission in infrared band.

  After all, we should acknowledge that current data are insufficient to draw a ultimate conclusion.
  At the resolution of IRAC, it is difficult to further identify
  the mechanism to produce nuclear infrared emission among different
  possibilities. Comprehensive study of the SED might provide more
  information on this issue.

  \section{Conclusion}
 We performed the Spitzer IRAC observations of 36 local elliptical galaxies.
 9 out of 36 galaxies display red core structure with nuclear infrared excess emission.
 The infrared excess emissions only and universally appear in galaxies with relatively luminous central
 AGN, strongly support a relation between the two. We also confirmed the correlation between the activities of AGN
 and optically observed circumnuclear dust and found this correlation also holds for extended dust in
 elliptical galaxies. We found correlation with considerable scatter between the luminosity of central
 AGN and excess emission, which indicates unresolved non-AGN contamination of excess emission in low-luminosity sources.
 While the specific mechanism to produce infrared emission could not be identified by current
 data, thermal origin from hot dust is supported by similar infrared
 spectral indices with Seyfert galaxies.
 In order to clarify origins for LLAGNs' infrared emission, a further study
 calls for multi-wavelength study, which will be provided in our
 next work.

 \section*{ACKNOWLEDGMENTS}
We are very grateful to Tom Jarrett and Feng Yuan for their valuable
advices. We are also thankful to Tao Wang and Song Huang for helpful
suggestions. This work is supported by Program for New Century
Excellent Talents in University (NCET), the national Natural Science
Foundation of China under grants 10878010 and 10633040, and the
National Basic Research Program (973 Program No. 2007CB815405). This
research has made use of NASA's Astrophysics Data System
Bibliographic Services and the NASA/IPAC Extragalactic Database
(NED), which is operated by the Jet Propulsion Laboratory,
California Institute of Technology, under contract with the National
Aeronautics and Space Administration. This work is based on
observations made with the Spitzer Space Telescope, which is
operated by the Jet Propulsion Laboratory, California Institute of
Technology, under NASA contract 1407.

\end{document}